\documentclass[12pt]{iopart}

\usepackage[utf8]{inputenc} 
\usepackage{float} 
\usepackage{enumitem}
\usepackage{bbold}
\usepackage[normalem]{ulem}
\usepackage{hyperref} 
\hypersetup{colorlinks=true, urlcolor=blue, citecolor=blue,linkcolor=blue}
\usepackage[all]{hypcap} 

\usepackage{graphicx}
\usepackage{dcolumn}
\usepackage{bm}
\usepackage{psfrag}
\usepackage{epsfig}
\usepackage{multirow}
\usepackage{color}
\usepackage{bbm}
\usepackage[FIGTOPCAP,raggedright,nooneline]{subfigure}
\usepackage{verbatim}
\usepackage{upgreek} 
\usepackage{tikz} 
\usepackage[mathscr]{euscript}

\makeatletter

\newcommand{\Rmnum}[1]{\expandafter\@slowromancap\romannumeral #1@}
\newcommand{\mi}{\mathrm{i}}
\newcommand{\jp}{\mathcal{P}}
\makeatother

\def\JH{{\cal H}}
\def\bra#1{\langle #1|}
\def\ket#1{ |#1 \rangle}

\def\JH{{\cal H}}
\def\Tr{\mbox{\rm Tr}}
\def\l{\ell}

\usepackage{iopams}  

\expandafter\let\csname equation*\endcsname\relax

\expandafter\let\csname endequation*\endcsname\relax

\usepackage{amsmath}

\begin{document}
\bibliographystyle{unsrt}

\title[]{Experimental Demonstration of a Quantum Controlled-SWAP Gate with Multiple Degrees of Freedom of a Single Photon}

\author{Feiran Wang$^{1}$, Shihao Ru$^{2}$, Yunlong Wang$^{2,*}$, Min An$^{2}$, Pei Zhang$^{2}$, Fuli Li$^{2}$}

\address{$^{1}$School of Science of Xi’an Polytechnic University, Xi'an 710048, China}
\address{$^{2}$Shaanxi Key Laboratory of Quantum Information and Quantum Optoelectronic Devices, School of Physics of Xi'an Jiaotong University, Xi'an 710049, China}
\address{$^{*}$Corresponding author: yunlong.wang@mail.xjtu.edu.cn}


\vspace{10pt}
\begin{indented}
\item[]February 2021
\end{indented}

\begin{abstract}
Optimizing the physical realization of quantum gates is important to build a quantum computer. The controlled-SWAP gate, also named Fredkin gate, can be widely applicable in various quantum information processing schemes.
In the present research, we propose and experimentally implement quantum Fredkin gate in a single-photon hybrid-degrees-of-freedom system.
Polarization is used as the control qubit, and SWAP operation is achieved in a four-dimensional Hilbert space spanned by photonic orbital angular momentum.
The effective conversion rate $\jp$ of the quantum Fredkin gate in our experiment is $(95.4\pm 2.6)\%$.
Besides, we find that a kind of Greenberger-Horne-Zeilinger-like states can be prepared by using our quantum Fredkin gate, and these nonseparale states can show its quantum contextual characteristic by the violation of Mermin inequality.
Our experimental design and coding method are useful for quantum computing and quantum fundamental study in high-dimensional and hybrid coding quantum systems.
 \end{abstract}

%
%
%
%
%

\section{Introduction}
Universal quantum computing can be realized with a series of basic quantum gates, including phase shift gate and Hadamard gate of single qubits, and controlled NOT gate of two qubits \cite{quantumgate1995,2005Cliffordgate,walther2005experimental,o2007optical,takeda2019toward}.
The multi-qubit gates, such as Fredkin (controlled SWAP) gate and Toffoli gate, can be decomposed by a set of single-qubit and two-qubit gates \cite{1996Fredkin,2013Toffoli}.
However, the inefficient synthesis of such multi-qubit gates increases the length and time scales of quantum system, and makes the gates further susceptible to their environment.
In other words, although basic quantum gates have been realized in many physical platforms and salient features of a quantum computer have been exhibited in proof-of-principle experiments \cite{kok2007linear,o2009photonic,ladd2010quantum}, difficulties of complex operations will emerge for scaling quantum systems.
The Fredkin gate is one of the typical examples.
Although there exist various theoretical proposals and experimental realizations of Fredkin gate in linear optics \cite{fiuravsek2006linear,gong2008methods,yu2015optimal,patel2016quantum,ono2017implementation,starek2018experimental,liu2020low}, those need post-selection strategies or are probabilistic. The corresponding quantum implementation is not simple enough.

Therefore, achieving multi-qubit gates with less decoherence and less error rate by utilizing as fewer quantum resources as possible is key to quantum information precessing.
There are mainly two tactics to solve it.
One is to exploit a system with accessible high dimensional states (qudits) in quantum information processing \cite{forbes2019quantum,erhard2020advances}.
A qudit can be seen as a quantum particle which is not limited to two levels but in principle can have any number of discrete levels.
Additionally, using a high-dimensional system to encode qubits or realize quantum logic gates, the interaction and decoherence between multi qubits would be more less.
Meanwhile qudits exhibit many merits such as enhancing channel capacities, the fault tolerance and noise resistant \cite{securitydlevel2002,lo2014secure,PRX2019noise}.
Another one is to encode qubits in multiple degrees of freedom (DoF) of a quantum system.
Even if extra DoFs, mostly, are used as auxiliary qubits, but can be directly used as computational qubits in some quantum information processing.

Furthermore, the orbital angular momentum (OAM), due to its high-dimensional quantum properties, various effective control and identification technologies, has attracted widespread attention \cite{rubinsztein2016roadmap,erhard2018twisted,shen2019optical,2019aqt}. From the perspective of quantum mechanics, the OAM occurs in discrete steps of $\ell\hbar$, where $\ell$ is an unbounded integer in principle \cite{allen1992orbital}. The basic characteristic of photons carrying OAM is the existence of the spiral phase factor $\exp(\mi\ell\phi)$, where $\phi$ is the azimuthal coordinate in the plane transverse to the propagation direction.
Besides, a wide range of research and review articles provides information on the generation, properties, as well as classical and quantum applications of multi-DoF entanglement (for example, OAMs coupled with polarization DoFs) \cite{karimi2015classical,doi:10.1116/1.5112027,doi:10.1116/5.0016007,paneru2020entanglement}.
Classical correlations are incapable of describing quantum correlations but proving to be useful in applications such as quantum metrology \cite{toppel2014classical}.
For single-quanta multi-DoF nonseparable states, they cannot lead to any quantum nonlocal correlations, but can be used to test other statistical models, such as noncontextual realistic models \cite{karimi2010spin,ru2020theoretical}.
In the present research, a kind of Greenberger-Horne-Zeilinger-like (GHZ-like) states generated by our Fredkin gate are used to test the quantum contextuality.

In this article, we focus on the experimental realization of deterministic quantum Fredkin gate with hybrid DoFs of a single photon. Polarization and OAM DoFs are used to span the Hilbert space $\JH_8=\JH_2^{\mathrm{p}}\otimes\JH_4^{\mathrm{o}}$, where the superscript o and p represent the orbital angular momentum and the polarization DoF, respectively, and the subscripts represent the corresponding Hilbert space dimension.
We use a specific set of quantum states as input states to test the performance of the Fredkin gate.
We utilize the conversion rate for verification of the gate, and all conversion rates measured are greater than $93\%$.  Meanwhile, we have quantum state tomography (QST) in Hilbert space $\JH_4^{\mathrm{o}}$ while the polarization of a given quantum state is horizontal, vertical or diagonal. In addition, the single-quanta nonseparable states can be generated by our quantum Fredkin gate, and we perform a Mermin inequality test to discuss quantum contextuality.
Our research paves the way for quantum computing and quantum fundamental study by using hybrid-DoF or high-dimension coding methods.

\section{Experiment setup}
\begin{figure}[!t]
\centering
\includegraphics[width=0.6\linewidth]{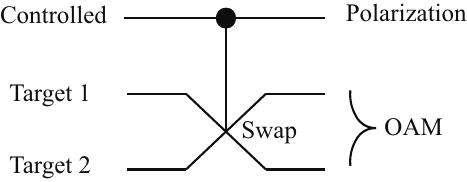}
\caption{The quantum Fredkin gate circuit. The control qubit is encoded by polarization DoF, and the two target qubits are encoded in a four-dimensional Hilbert space spanned by OAM quantum numbers of $-2,-1,0$ and $1$.}\label{fig:circuit}
\end{figure}

The Fredkin gate has been experimentally implemented with linear optics on multi-photon single-DoF system \cite{patel2016quantum}.
But something more meaningfully, coding in a single photon with hybrid DoFs to realize Fredkin gate is much more economical than a multi-photon single-DoF coding method.
As will be described below, our experimental proposal is deterministic and there is no photon loss in theory.
The Fredkin gate can be written as
\begin{align*}
U_{\mathrm{Fred}}=\ket{0}\bra{0}\otimes I_4+\ket{1}\bra{1}\otimes U_{\mathrm{SWAP}}.
\end{align*}
As shown in Fig.~\ref{fig:circuit}, it is a three-qubit controlled-swap gate, which makes the states of the two target qubits swapped according to the condition of the control qubit. The control qubit is encoded by polarization DoF ($\ket{V}\to\ket{0}$ and $\ket{H}\to\ket{1}$), and the two target qubits are encoded in a four-dimensional Hilbert space spanned by OAM states from $\ket{\ell=-2}$ to $\ket{\ell=+1}$, of which specific coding methods are $\ket{\ell=-1}\to\ket{00}$, $\ket{\ell=-2}\to\ket{01}$, $\ket{\ell=0}\to\ket{10}$ and $\ket{\ell=+1}\to\ket{11}$.

\begin{figure}[!t]
\centering
\includegraphics[width=0.6\linewidth]{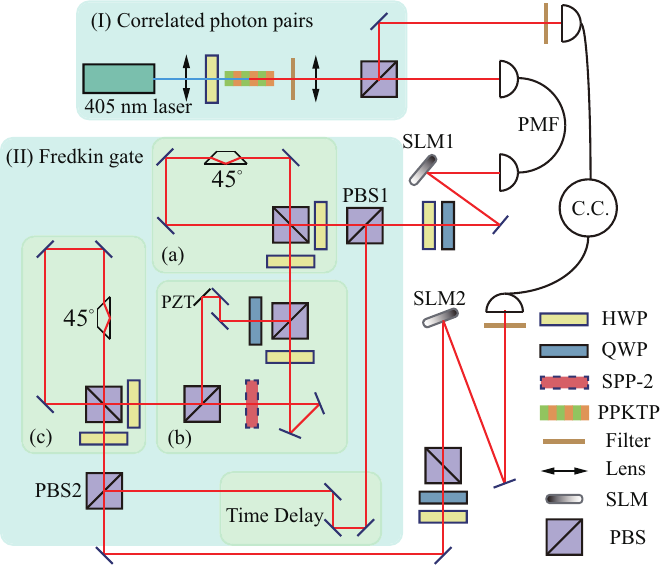}
\caption{Quantum Fredkin gate for three qubits encoded into SAM and OAM DoFs of a single photon. A continuous wave laser with average power of $12$ mW at $405$ nm pumps a nonlinear crystal of PPKTP cut for degenerate type \Rmnum{2} collinear phase matching which emits correlated photon pairs at $810$ nm. Photons from the source are spatially filtered to the fundamental Gaussian mode using a PMF. The vertical photon from correlated photon source is directly coupled into PMF and detected by SPAD as a trigger signal.
SLM1, a reflector and a set of plates are used to separately modulate traverse and polarization modes of another photon.
Part (\Rmnum{2}) is to achieve the quantum Fredkin gate here, in which Figs. (a)-(c) is to achieve SWAP operation in OAM-encoding Hilbert space.
The combination of QWP, HWP and PBS is to modulate the polarization modes of photons to the horizontal polarization, which is responded by SLM2, and the tomography of OAM-encoding Hilbert space $\JH_4^{\mathrm{o}}$ is performed with SLM2.
The phase of this photon is flattened, coupled into PMF and detected by another SPAD.
PPKTP: periodically poled potassium titanyl phosphate; HWP: half wave plate; QWP: quarter wave plate; PBS: polarizing beam splitter; SLM: spatial light modulator; SPP: spiral phase plate; PMF: polarization maintaining single mode fiber; SPAD: single-photon avalanche detector.}\label{fig:exp}
\end{figure}

Figure~\ref{fig:exp} illustrates our linear optical scheme for the quantum Fredkin gate.
The four-dimensional OAM space is from $\ket{\ell=-2}$ to $\ket{\ell=+1}$ here, and any arbitrary superposition OAM states can be generated by spatial light modulator (SLM).
The half wave plate (HWP) and quarter wave plate (QWP) positioned behind SLM1 are used to generate an arbitrary polarization quantum state for the control qubit.
The core component of our experimental scheme is an altered Mach-Zehnder interferometer composed of Figs. \ref{fig:exp}(a)-(c), time delay, and two polarized beamer splitters (PBSs) before and after them in the part (\Rmnum{2}) of Fig.~\ref{fig:exp}.
PBS 1 and 2 are used as controlled operation for polarization DoF.
The computational state $\ket{1}$ of the control qubit corresponds to the horizontal polarization photon propagating in the upper interferometer arm, while the state $\ket{0}$ is represented by vertical polarization photon propagating in the lower arm of the polarization Mach-Zehnder interferometer.
The time-delay part in the lower arm is used to balance the path difference between two arms.

Notice that the Figs. \ref{fig:exp}(a) and \ref{fig:exp}(c) are identity in the experimental settings.
All the angles of HWPs are at $\pi/8$ and $3\pi/8$, sequentially, and the angles of Dove primes are at $\pi/4$. In general, if the Dove Prism is rotated by an angle of $\alpha$, the OAM mode can be operated in the following manner,
\begin{equation}
\ket{\l}\xrightarrow[]{\mathrm{Dove}} \mi \exp(\mi 2 \ell \alpha)\left|-\l\right>.
\end{equation}
While the forward propagating photon in an OAM state $\ket{\l}$ would be added a phase of $\exp(\mi 2 \ell \alpha)$, the backward propagating photon carrying the same OAM mode would be added an opposite phase of $\exp(-\mi 2 \ell \alpha)$.
The embedded Sagnac interferometers (the Figs. \ref{fig:exp}(a) and \ref{fig:exp}(c)) label as parity sorters, and are designed to distinguish photons based on the parity of its OAM quantum numbers. Strictly speaking, this kind of parity sorter can be used to constitute quantum non-demolition measurement in hyperentangled Bell states measurement.

The Fig. \ref{fig:exp}(b) is a Mach-Zehnder interferometer in which odd spatial mode photons propagate in upper arm and even spatial mode propagate in lower arm. It is to swap the even modes $\ket{\ell=0}$ and $\ket{\ell=-2}$ themselves by using a spiral phase plate (SPP).
We also add a QWP at $0^{\circ}$ and a HWP at $\pi/2$ in Fig. \ref{fig:exp}(b) to compensate the phase induced by the overall reflection number on the relative phase between odd and even modes.
A PZT in the upper arm combined with a motorized translation stage is to ensure the optical path of two arm same. That is, no additional phase is introduced. The contrast of each interferometer in our experiment is between 94\% and 95\%.
Based on the above modules, we can experimentally implement the Fredkin gate with hybrid DoFs.

In order to explain more clearly, assume that the OAM initial state is generated by SLM1 as follows,
\begin{align}
\ket{\psi_1}=\ket{H}\otimes\left(C_{-2}\ket{-2}+C_{-1}\ket{-1}+C_0\ket{0}+C_{+1}\ket{+1}\right),
\end{align}
where $\sum_{i=-2}^{1}\left|C_i\right|^2=1$.
After a reflector, a QWP and a HWP, it becomes
\begin{align}
\ket{\psi_2}&=\mi\left(\cos\alpha\ket{H}+e^{\mi\beta}\sin\alpha\ket{V}\right)\notag\\
&\otimes\left(C_{-2}\ket{+2}+C_{-1}\ket{+1}+C_0\ket{0}+C_{+1}\ket{-1}\right).
\end{align}
Here we consider the situation of the upper and the lower arms, respectively. Only some mirrors are included in the lower arm for a time delay. Therefore, the quantum state of photon propagating through the lower arm can be written as
\begin{align}\label{lower}
\ket{\psi_d}&=-e^{\mi\beta}\sin\alpha\ket{V}\notag\\
&\otimes\left(C_{-2}\ket{-2}+C_{-1}\ket{-1}+C_0\ket{0}+C_{+1}\ket{+1}\right)
\end{align}
after PBS2.
For the upper arm, we consider that the four different OAM spatial modes transform through Figs. \ref{fig:exp}(a), (b), and (c).
As mentioned above, the quantum state is transferred into
\begin{align}\label{upper}
\ket{\psi_u}&=-\cos\alpha\ket{H}\notag\\
&\otimes\left(C_{-2}\ket{0}+C_{-1}\ket{-1}+C_0\ket{-2}+C_{+1}\ket{+1}\right).
\end{align}
Comparing Eq.~(\ref{lower}) with Eq.~(\ref{upper}), one can see that no extra relative phase of quantum states is added between the two arms and the global phase can be ignored. Besides, recall that our coding methods of four-dimensional OAM Hilbert space are $\ket{\l=-1}\to\ket{00}$, $\ket{\l=-2}\to\ket{01}$, $\ket{\l=0}\to\ket{10}$ and $\ket{\l=+1}\to\ket{11}$, that is, two qubits can be encoded with a four-dimensional OAM qudit, $\JH_4^{\mathrm{o}}=:\JH_2\otimes\JH_2$.
Therefore the final quantum state becomes
\begin{align}
\ket{\psi_f}&=\ket{\psi_u}+\ket{\psi_d}\notag\\
&=\left(\ket{0}\bra{0}\otimes I_4+\ket{1}\bra{1}\otimes U_{\mathrm{SWAP}}\right)\ket{\psi_2}\notag\\
&=U_{\mathrm{Fred}}\ket{\psi_2},
\end{align}
which satisfies the quantum Fredkin gate.

\section{Measurement scheme and experimental results}
For measurements, we adopt the crosstalk measurements between the input and output states in the computational basis and perform the QST for the output states in the OAM Hilbert space.
The computational basis includes the eight eigenstates (from $\ket{000}$ to $\ket{111}$) of $\sigma_z$.
Experimentally, a set of wave plates and a SLM can achieve computational basis measurements in polarization and OAM hybrid DoFs, and this setup can also be used for QST.
\begin{figure}[!b]
\centering
\subfigure[]{
\begin{minipage}[b]{0.7\textwidth}
\includegraphics[width=1\textwidth]{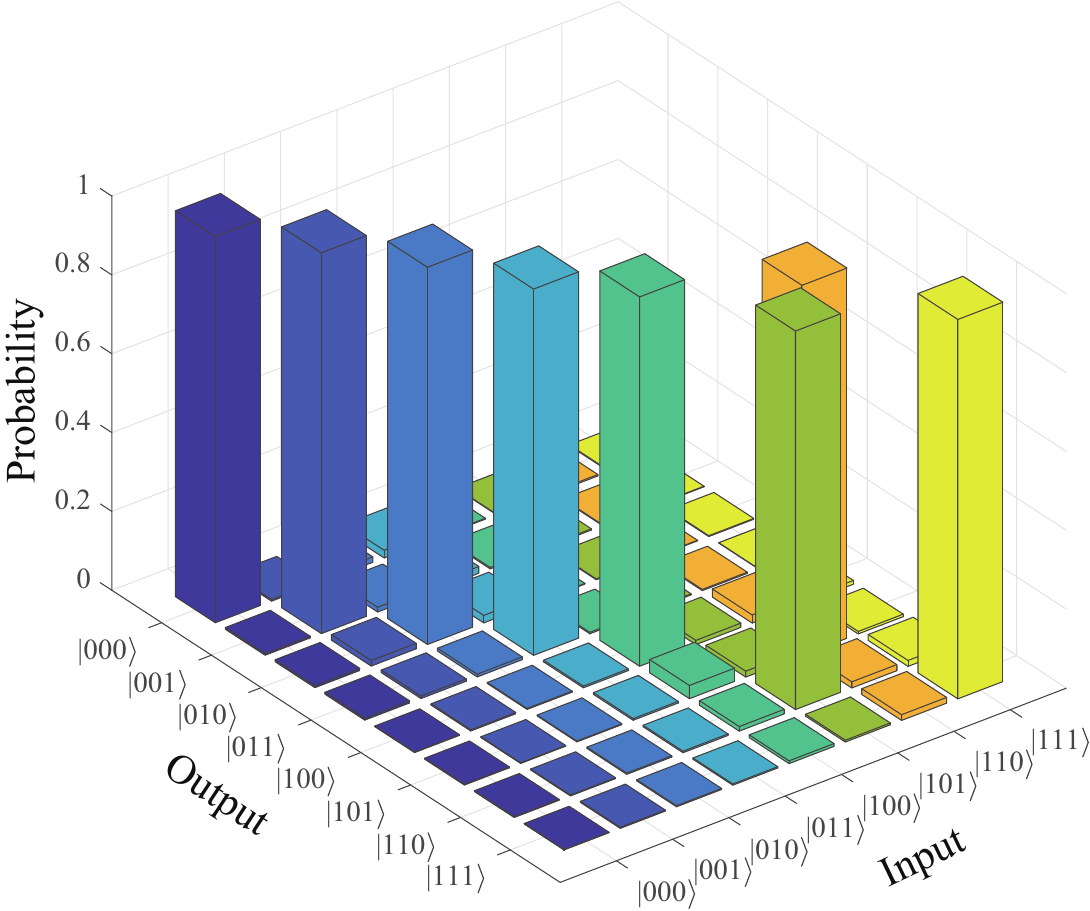}
\end{minipage}
}
\subfigure[]{
\begin{minipage}[b]{0.7\textwidth}
\includegraphics[width=1\textwidth]{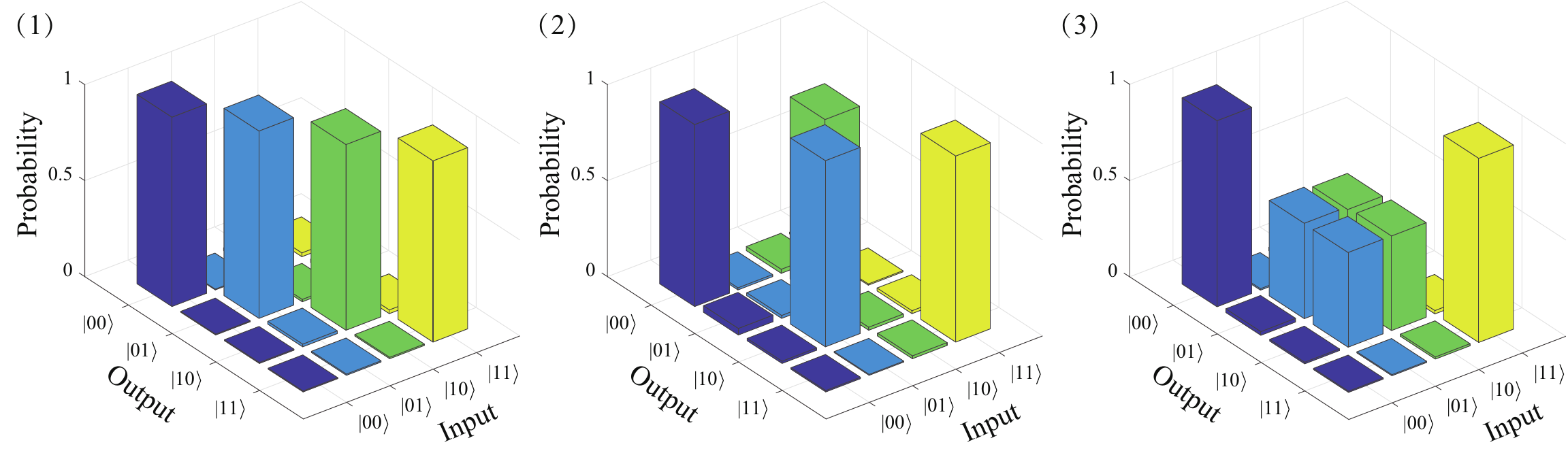}
\end{minipage}
}
\caption{Truth tables of the Fredkin gate. (a) After preparing three qubits in one of the eight basis input states from $\ket{000}$ to $\ket{111}$, the probabilities of all basis output states are measured $1$ minute. (b) For supplementary explanation, the subgraphs are the truth tables corresponding to the target qubits while the state of control qubit is $\ket{0}$, $\ket{1}$ and $(\ket{0}+\ket{1})/\sqrt{2}$, respectively.}
\label{fig:cross}
\end{figure}
The final maximum coincidence counts for correlated photon pairs are 58.56 kHz in average when the pump power is 12 mW, the accidental counts are only 5 Hz, and the coincidence efficiency is about 11.6\%. The single-photon detectors used in our experiment are all SPCM-AQRH-14-FC, EXCELITAS. Its single-photon efficiency at 810 nm is about 62\%. The time delay between correlated photon pair is 8.5 ns in our experimental setup.

To exhibit the quality of the Fredkin gate, we define the effective conversion rate $\jp$ for the transformation, according to $\jp(i,j)=N_{ij}/\sum_{mn}N_{mn}$, where $N_{ij}$ corresponds to the coincidence count entries in the diagonal crosstalk matrix.
It is obtained from the crosstalk measurements between the input and output modes.

The truth table depicted in Fig.~\ref{fig:cross}(a) shows the probabilities of all computational basis states after applying the Fredkin gate to each of the computational basis states.
The effective conversion rate of the truth table is $\jp=(95.4\pm 2.6)\%$.
For more detailed classification, we measure the outputs of the gate for each of the four possible computational basis input states ($\ket{00}$, $\ket{01}$, $\ket{10}$ and $\ket{11}$) while control qubit 1 is being in the different states.
If the state of control qubit is $\ket{0}$, Fig.~\ref{fig:cross}-(b1) shows that the states of target qubits $2$ and $3$ indeed invariant, satisfying the theoretical prediction.
Figure~\ref{fig:cross}-(b2) shows that the unitary transformation of qubits $2$ and $3$ actually is a SWAP gate if control qubit is in the state $\ket{1}$.
The effective conversion rate of this SWAP gate is $\jp_\mathrm{SWAP}=95.9\%$.
These reveal the characteristic properties of the Fredkin gate, named that a SWAP operation is applied on the two target qubits if the control qubit is in the state $\ket{1}$.
In addition, in order to manifest well coherence of our experimental setup, we consider the situation where
the control qubit is in the  state $(\ket{0}+\ket{1})/\sqrt{2}$ and the gate is in a superposition of the SWAP and identity operations. The results are shown in Fig.~\ref{fig:cross}-(b3).

\begin{figure}[!t]
\centering
\includegraphics[width=0.9\linewidth]{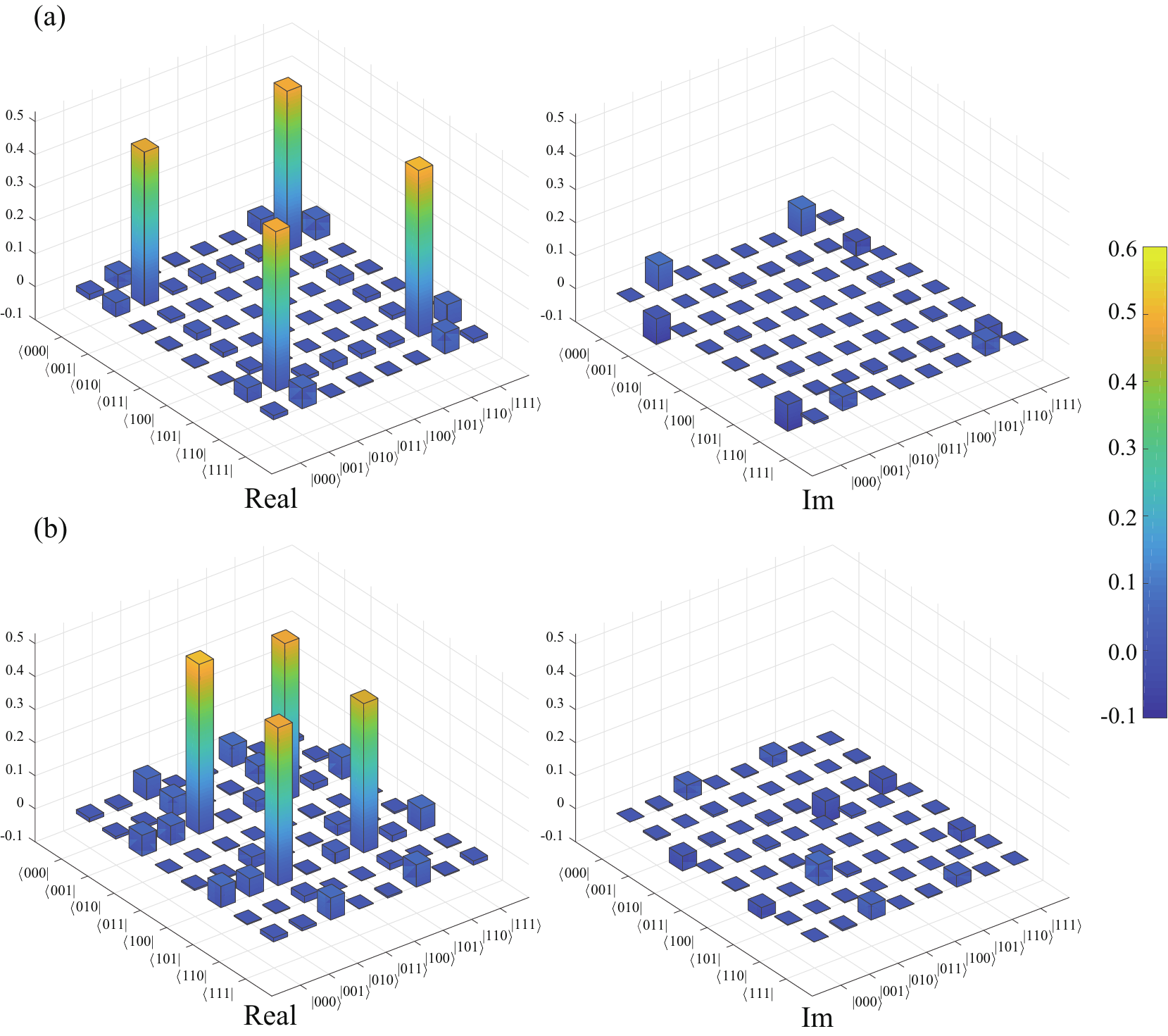}
\caption{The reconstructed density matrices for two nonseparable states.
The control and two target qubits are measured in the $D/A$ basis ($\sigma_x$); in the $R/L$ basis ($\sigma_y$); and in the $H/V$ basis ($\sigma_z$).
Part (a) is $\ket{\ket{\psi_{001}}}$ and part (b) is $\ket{\psi_{010}}$.}
\label{fig:tomo}
\end{figure}

Furthermore, using our Fredkin gate, one can produce maximally entangled three-qubit states, GHZ-like states $\ket{\psi _{\mu \lambda \omega}}=\sum\limits_{j = 0,1}{(-1)^{\mu j}\ket{j,j \oplus \lambda ,j \oplus \omega}/{\sqrt 2}} $, in which $\mu,\lambda,\omega, j$ vary 0 or 1.
Two nonseparable states  $\ket{\psi_{001}}$ and $\ket{\psi_{010}}$ are illustrated in Fig.~\ref{fig:tomo}, namely,
\begin{align*}
&\left(\ket{0}+\ket{1}\right)\ket{01}/\sqrt{2}\xrightarrow[]{\mathrm{Fredkin}}\ket{\psi_{001}}:\left(\ket{001}+\ket{110}\right)/\sqrt{2},\\
&\left(\ket{0}+\ket{1}\right)\ket{10}/\sqrt{2}\xrightarrow[]{\mathrm{Fredkin}}\ket{\psi_{010}}:\left(\ket{010}+\ket{101}\right)/\sqrt{2}.
\end{align*}
A full state reconstruction can be carried out by a set of $27$ measurements setting ($\sigma_x\sigma_x\sigma_x,\sigma_x\sigma_x\sigma_y,\cdots,\sigma_z\sigma_z\sigma_z$), effectively resulting in an overcomplete set of $216$ projective measurements.
Figure~\ref{fig:tomo} shows the real and imagine parts of the reconstructed density matrices of these two nonseparable states, which we used a maximum-likelihood algorithm to get.
We measure the fidelity between the output state $\varrho_{o}$ and the desired state $\varrho_{e}$ as $F(\varrho_o,\varrho_e)=\left(\Tr\sqrt{\sqrt{\varrho_o}\varrho_e\sqrt{\varrho_o}}\right)^2$, and obtain $F=(96.8\pm2.3)\%$ for $\ket{\psi_{001}}$ and $F=(96.0\pm1.7)\%$ for $\ket{\psi_{010}}$.
The errors are calculated using Monte Carlo simulations from our measurement sampling.

Through performing further measurements on the nonseparable state, we can characterize quantum contextuality for demonstrating the nonclassical nature of quantum mechanics \cite{karimi2010spin,ru2020theoretical}.
GHZ states can be used to show a strong contradiction between the noncontextual hidden variable theories and quantum mechanics. As we discussed in Introduction, such single-quanta nonseparable state cannot be used to verify quantum nonlocality, but can be used to verify quantum contextuality by the violation of Mermin inequality.
The Mermin inequality can be written as
\begin{align}
S_M=\left|\left<\sigma_x\sigma_x\cdot \sigma_x\right>+\left<\sigma_x\sigma_y\cdot \sigma_y\right>+\left<\sigma_y\sigma_x\cdot \sigma_y\right>-\left<\sigma_y\sigma_y\cdot \sigma_x\right>\right|\le2,
\end{align}
\begin{table}[!b]
\centering
\caption {The correlation function values of $\ket{\psi_{001}}$.}
\label{tab:tab}
\begin{tabular}{cccc}
\hline\hline
$\left<\sigma_x\sigma_x\cdot\sigma_x\right>$&$\left<\sigma_x\sigma_y\cdot\sigma_y\right>$&$\left<\sigma_y\sigma_x\cdot\sigma_y\right>$&$\left<\sigma_y\sigma_y\cdot\sigma_x\right>$\\
\hline
$0.970\pm0.014$&$0.942\pm0.010$&$0.944\pm0.016$&$-0.962\pm0.011$\\
\hline\hline
\end{tabular}
\end{table}
which holds for noncontextual hidden variable theories since each of the six variables are assigned to predefined values $+1$ or $-1$.
However, quantum mechanics predicts the maximal value of the computation to be 4 (with ideal equipments).
We calculate it from the QST results of $\ket{\psi_{001}}$, the corresponding correlation functions are listed in TABLE~\ref{tab:tab},  and the expectation value of $S_M$ in our experiment can be obtained as $3.818\pm0.016$.
This value is greater than the maximum predicted under the noncontextual hidden variable theories background.
Therefore, it actually discards that noncontextual hidden variable theories are extensions of quantum mechanics.
All those results show that we have successfully realized the quantum Fredkin gate, and the feasibility of manipulating multiple DoFs simultaneously.

\section{Conclusion}
In summary, an efficient method to construct quantum Fredkin gate with hybrid DoFs of a single photon is presented in our study. Using the polarization DoFs and the four-dimensional OAM space to encode the control qubit and the two target qubits, respectively, we implement a deterministic optical Fredkin gate without photon loss.
This coding method can also be extended to $n$-qubit Fredkin gate, and can be used to implement {\small CNOT} gate and Toffoli gate.
The gate demonstrated here uses the ability to sort even and odd parity OAM modes as a basic building block. This concept can be extended to other photonic degrees of freedom such as path and frequency. The photons with odd modes experience an extra reflection which corresponds to the NOT gate operation for OAM modes. In addition, the spiral phase plate has achieved a mode increase for even modes. The above transformations, such as parity sorter, NOT gate etc., are usually used in quantum computing, which can be implemented not only in OAM DoF, but also in other quantum systems.  Therefore, our method may also be used in other proposed quantum systems such as trapped ions \cite{muthukrishnan2000multivalued,klimov2003qutrit}, cold atoms \cite{smith2013quantum}, and superconducting circuits \cite{hofheinz2009synthesizing} for constructing similar quantum logic gates.
\vspace{1cm}

\section*{Acknowledgments}
This work was in part supported by the National Natural Science Foundation of China (Grant Nos. 11534008, 11804271, 91736104 and 12074307), Ministry of Science and Technology of China (2016YFA0301404) and China Postdoctoral Science Foundation via Project No. 2020M673366.

\vspace{1cm}

\section*{References}


\begin{thebibliography}{10}

\bibitem{quantumgate1995}
Adriano Barenco, Charles~H. Bennett, Richard Cleve, David~P. DiVincenzo, Norman
  Margolus, Peter Shor, Tycho Sleator, John~A. Smolin, and Harald Weinfurter.
\newblock Elementary gates for quantum computation.
\newblock {\em Phys. Rev. A}, 52:3457--3467, Nov 1995.

\bibitem{2005Cliffordgate}
Sergey Bravyi and Alexei Kitaev.
\newblock Universal quantum computation with ideal clifford gates and noisy
  ancillas.
\newblock {\em Phys. Rev. A}, 71:022316, Feb 2005.

\bibitem{walther2005experimental}
Philip Walther, Kevin~J Resch, Terry Rudolph, Emmanuel Schenck, Harald
  Weinfurter, Vlatko Vedral, Markus Aspelmeyer, and Anton Zeilinger.
\newblock Experimental one-way quantum computing.
\newblock {\em Nature}, 434(7030):169--176, 2005.

\bibitem{o2007optical}
Jeremy~L O'brien.
\newblock Optical quantum computing.
\newblock {\em Science}, 318(5856):1567--1570, 2007.

\bibitem{takeda2019toward}
S~Takeda and A~Furusawa.
\newblock Toward large-scale fault-tolerant universal photonic quantum
  computing.
\newblock {\em APL Photonics}, 4(6):060902, 2019.

\bibitem{1996Fredkin}
John~A. Smolin and David~P. DiVincenzo.
\newblock Five two-bit quantum gates are sufficient to implement the quantum
  fredkin gate.
\newblock {\em Phys. Rev. A}, 53:2855--2856, Apr 1996.

\bibitem{2013Toffoli}
Nengkun Yu, Runyao Duan, and Mingsheng Ying.
\newblock Five two-qubit gates are necessary for implementing the toffoli gate.
\newblock {\em Phys. Rev. A}, 88:010304, Jul 2013.

\bibitem{kok2007linear}
Pieter Kok, William~J Munro, Kae Nemoto, Timothy~C Ralph, Jonathan~P Dowling,
  and Gerard~J Milburn.
\newblock Linear optical quantum computing with photonic qubits.
\newblock {\em Rev. Mod. Phys.}, 79(1):135, 2007.

\bibitem{o2009photonic}
Jeremy~L O'brien, Akira Furusawa, and Jelena Vu{\v{c}}kovi{\'c}.
\newblock Photonic quantum technologies.
\newblock {\em Nat. Photonics}, 3(12):687--695, 2009.

\bibitem{ladd2010quantum}
Thaddeus~D Ladd, Fedor Jelezko, Raymond Laflamme, Yasunobu Nakamura,
  Christopher Monroe, and Jeremy~Lloyd O’Brien.
\newblock Quantum computers.
\newblock {\em Nature}, 464(7285):45--53, 2010.

\bibitem{fiuravsek2006linear}
Jarom{\'\i}r Fiur{\'a}{\v{s}}ek.
\newblock Linear-optics quantum toffoli and fredkin gates.
\newblock {\em Phys. Rev. A}, 73(6):062313, 2006.

\bibitem{gong2008methods}
Yan-Xiao Gong, Guang-Can Guo, and Timothy~C Ralph.
\newblock Methods for a linear optical quantum fredkin gate.
\newblock {\em Phys. Rev. A}, 78(1):012305, 2008.

\bibitem{yu2015optimal}
Nengkun Yu and Mingsheng Ying.
\newblock Optimal simulation of deutsch gates and the fredkin gate.
\newblock {\em Phys. Rev. A}, 91(3):032302, 2015.

\bibitem{patel2016quantum}
Raj~B Patel, Joseph Ho, Franck Ferreyrol, Timothy~C Ralph, and Geoff~J Pryde.
\newblock A quantum fredkin gate.
\newblock {\em Sci. Adv.}, 2(3):e1501531, 2016.

\bibitem{ono2017implementation}
Takafumi Ono, Ryo Okamoto, Masato Tanida, Holger~F Hofmann, and Shigeki
  Takeuchi.
\newblock Implementation of a quantum controlled-swap gate with photonic
  circuits.
\newblock {\em Sci. Rep.}, 7(1):1--9, 2017.

\bibitem{starek2018experimental}
Robert St{\'a}rek, Martina Mikov{\'a}, Ivo Straka, Miloslav Du{\v{s}}ek,
  Miroslav Je{\v{z}}ek, Jarom{\'\i}r Fiur{\'a}{\v{s}}ek, and Michal
  Mi{\v{c}}uda.
\newblock Experimental realization of swap operation on hyper-encoded qubits.
\newblock {\em Opt. Express}, 26(7):8443--8452, 2018.

\bibitem{liu2020low}
Wen-Qiang Liu, Hai-Rui Wei, and Leong-Chuan Kwek.
\newblock Low-cost fredkin gate with auxiliary space.
\newblock {\em Phys. Rev. Appl.}, 14(5):054057, 2020.

\bibitem{forbes2019quantum}
Andrew Forbes and Isaac Nape.
\newblock Quantum mechanics with patterns of light: Progress in high
  dimensional and multidimensional entanglement with structured light.
\newblock {\em AVS Quantum Sci.}, 1(1):011701, 2019.

\bibitem{erhard2020advances}
Manuel Erhard, Mario Krenn, and Anton Zeilinger.
\newblock Advances in high-dimensional quantum entanglement.
\newblock {\em Nat. Rev. Phys.}, pages 1--17, 2020.

\bibitem{securitydlevel2002}
Nicolas~J. Cerf, Mohamed Bourennane, Anders Karlsson, and Nicolas Gisin.
\newblock Security of quantum key distribution using $\mathit{d}$-level
  systems.
\newblock {\em Phys. Rev. Lett.}, 88:127902, Mar 2002.

\bibitem{lo2014secure}
Hoi-Kwong Lo, Marcos Curty, and Kiyoshi Tamaki.
\newblock Secure quantum key distribution.
\newblock {\em Nat. Photonics}, 8(8):595--604, 2014.

\bibitem{PRX2019noise}
Sebastian Ecker, Fr\'ed\'eric Bouchard, Lukas Bulla, Florian Brandt, Oskar
  Kohout, Fabian Steinlechner, Robert Fickler, Mehul Malik, Yelena Guryanova,
  Rupert Ursin, and Marcus Huber.
\newblock Overcoming noise in entanglement distribution.
\newblock {\em Phys. Rev. X}, 9:041042, Nov 2019.

\bibitem{rubinsztein2016roadmap}
Halina Rubinsztein-Dunlop, Andrew Forbes, Michael~V Berry, Mark~R Dennis,
  David~L Andrews, Masud Mansuripur, Cornelia Denz, Christina Alpmann, Peter
  Banzer, Thomas Bauer, et~al.
\newblock Roadmap on structured light.
\newblock {\em J. Opt.}, 19(1):013001, 2016.

\bibitem{erhard2018twisted}
Manuel Erhard, Robert Fickler, Mario Krenn, and Anton Zeilinger.
\newblock Twisted photons: new quantum perspectives in high dimensions.
\newblock {\em Light: Sci. Appl.}, 7(3):17146--17146, 2018.

\bibitem{shen2019optical}
Yijie Shen, Xuejiao Wang, Zhenwei Xie, Changjun Min, Xing Fu, Qiang Liu, Mali
  Gong, and Xiaocong Yuan.
\newblock Optical vortices 30 years on: Oam manipulation from topological
  charge to multiple singularities.
\newblock {\em Light: Sci. Appl.}, 8(1):1--29, 2019.

\bibitem{2019aqt}
Daniele Cozzolino, Beatrice Da~Lio, Davide Bacco, and Leif~Katsuo Oxenlowe.
\newblock High-dimensional quantum communication: Benefits, progress, and
  future challenges.
\newblock {\em Adv. Quantum Technol.}, 2(12):1970073, 2019.

\bibitem{allen1992orbital}
Les Allen, Marco~W Beijersbergen, RJC Spreeuw, and JP~Woerdman.
\newblock Orbital angular momentum of light and the transformation of
  laguerre-gaussian laser modes.
\newblock {\em Phys.Rev. A}, 45(11):8185, 1992.

\bibitem{karimi2015classical}
Ebrahim Karimi and Robert~W Boyd.
\newblock Classical entanglement?
\newblock {\em Science}, 350(6265):1172--1173, 2015.

\bibitem{doi:10.1116/1.5112027}
Andrew Forbes and Isaac Nape.
\newblock Quantum mechanics with patterns of light: Progress in high
  dimensional and multidimensional entanglement with structured light.
\newblock {\em AVS Quantum Sci.}, 1(1):011701, 2019.

\bibitem{doi:10.1116/5.0016007}
Jinwen Wang, Francesco Castellucci, and Sonja Franke-Arnold.
\newblock Vectorial light–matter interaction: Exploring spatially structured
  complex light fields.
\newblock {\em AVS Quantum Sci.}, 2(3):031702, 2020.

\bibitem{paneru2020entanglement}
Dilip Paneru, Eliahu Cohen, Robert Fickler, Robert~W Boyd, and Ebrahim Karimi.
\newblock Entanglement: quantum or classical?
\newblock {\em Rep. Prog. Phys.}, 83(6):064001, 2020.

\bibitem{toppel2014classical}
Falk T{\"o}ppel, Andrea Aiello, Christoph Marquardt, Elisabeth Giacobino, and
  Gerd Leuchs.
\newblock Classical entanglement in polarization metrology.
\newblock {\em New J. Phys.}, 16(7):073019, 2014.

\bibitem{karimi2010spin}
Ebrahim Karimi, Jonathan Leach, Sergei Slussarenko, Bruno Piccirillo, Lorenzo
  Marrucci, Lixiang Chen, Weilong She, Sonja Franke-Arnold, Miles~J Padgett,
  and Enrico Santamato.
\newblock Spin-orbit hybrid entanglement of photons and quantum contextuality.
\newblock {\em Phys. Rev. A}, 82(2):022115, 2010.

\bibitem{ru2020theoretical}
Shihao Ru, Weidong Tang, Yunlong Wang, Feiran Wang, Pei Zhang, and Fuli Li.
\newblock Theoretical and experimental investigation of gauge-like equivalent
  quantum noncontextual inequalities.
\newblock {\em arXiv preprint arXiv:2010.05266}, 2020.

\bibitem{muthukrishnan2000multivalued}
Ashok Muthukrishnan and Carlos~R Stroud~Jr.
\newblock Multivalued logic gates for quantum computation.
\newblock {\em Phys. Rev. A}, 62(5):052309, 2000.

\bibitem{klimov2003qutrit}
AB~Klimov, R~Guzman, JC~Retamal, and Carlos Saavedra.
\newblock Qutrit quantum computer with trapped ions.
\newblock {\em Phys. Rev. A}, 67(6):062313, 2003.

\bibitem{smith2013quantum}
Aaron Smith, Brian~E Anderson, Hector Sosa-Martinez, Carlos~A Riofrio, Ivan~H
  Deutsch, and Poul~S Jessen.
\newblock Quantum control in the cs 6 s 1/2 ground manifold using
  radio-frequency and microwave magnetic fields.
\newblock {\em Phys. Rev. Lett.}, 111(17):170502, 2013.

\bibitem{hofheinz2009synthesizing}
Max Hofheinz, H~Wang, Markus Ansmann, Radoslaw~C Bialczak, Erik Lucero, Matthew
  Neeley, AD~O'connell, Daniel Sank, J~Wenner, John~M Martinis, et~al.
\newblock Synthesizing arbitrary quantum states in a superconducting resonator.
\newblock {\em Nature}, 459(7246):546--549, 2009.

\end{thebibliography}

\end{document}